\documentclass[english]{revtex4}
\usepackage[T1]{fontenc}
\usepackage[latin9]{inputenc}
\usepackage{amsmath}
\usepackage{amssymb}
\usepackage{graphicx}
\usepackage{esint}

\makeatletter
\@ifundefined{textcolor}{}
{%
 \definecolor{BLACK}{gray}{0}
 \definecolor{WHITE}{gray}{1}
 \definecolor{RED}{rgb}{1,0,0}
 \definecolor{GREEN}{rgb}{0,1,0}
 \definecolor{BLUE}{rgb}{0,0,1}
 \definecolor{CYAN}{cmyk}{1,0,0,0}
 \definecolor{MAGENTA}{cmyk}{0,1,0,0}
 \definecolor{YELLOW}{cmyk}{0,0,1,0}
}

\usepackage{babel}

\makeatother

\usepackage{babel}
\begin{document}

\title{Macrodeterminism without non-invasive measurability}

\author{Ramon Lapiedra}

\affiliation{Departament d'Astronomia i Astrofísica, Universitat de València,
46100 Burjassot, València, Spain\\
 }

\author{A. Pérez$^{1}$}

\affiliation{Departament de Física Teòrica and IFIC, Universitat de València-CSIC,
46100-Burjassot, València, Spain.\\
 }
\begin{abstract}
We propose a definition of determinism for a physical system that
includes, besides the measured system, the measurement device and
the surrounding environment. This \textit{enlarged system} is assumed
to follow a predefined trajectory starting from some (unknown) initial
conditions that play the role of hidden variables for the experiment.
These assumptions, which are different from realism, allow us to derive
some Leggett-Garg inequalities, which are violated by Quantum Mechanics
in the particular case of consecutive measurements on individual photon
polarizations. 
\end{abstract}

\pacs{04.20.-q, 98.80.Jk}

\maketitle
$^{1}$Tel. Nr. (34) 96-354-4551, Fax (34)-96-354-3381, e-mail: Armando.Perez@uv.es

\section{Introduction}

Using the terminology of \cite{0953-8984-14-15-201}, macroscopic
realism or macrorealism for a system entails the following three postulates:

(1) \emph{Macrorealism per se}: A macroscopic object which has available
to it two or more macroscopically distinct states is at any given
time in a definite one of those states. (2) \emph{Non-invasive measurability}
(NIM): Performing a measurement on the system has no effect on the
capacity of predicting ulterior measurement outcomes. (3)\emph{ Induction}:
The properties of the system are determined exclusively by the initial
conditions and, in particular, not by final conditions.

It is well known that these assumptions allow us to prove the Leggett-Garg
inequalities for successive measurements \cite{PhysRevLett.54.857}.
A violation of these inequalities leads to the macrorealism failure,
through the invalidation of at least one of the above assumptions.
Furthermore, in \cite{PhysRevA.87.052115} it is pointed out that
macrorealism also leads to a new property to add to the previous ones,
the so-called \emph{no-signaling in time}, which means that 'a measurement
does not change the outcome statistics of a later measurement'. The
authors of \cite{PhysRevA.87.052115} stress the independence of this
third assumption from the other ones.

A system made of a photon, whose polarization is consecutively measured
along different directions, randomly chosen among three selected and
fixed directions, can be seen, in accordance with Quantum Mechanics
(QM), to violate the no-signaling in time condition and also (see
for instance Section III) the Leggett-Garg inequalities. This second
violation had to be expected since the above NIM condition (2) is
by no means satisfied in the present case. But what about assumption
(1), macrorealism \emph{per se}, has it also to be invalidated in
this case? Instead of trying to answer this question, we will recast
it in the following one: Do the successive measurement outcomes obey
determinism in the specific way this term is used in classical physics,
say Laplacian determinism, where the evolution of the system is dictated
by some initial conditions? At first sight, the question seems nonsense
since the measured photon is not an isolated system: The measurement
device acts on the measured system, after randomly choosing different
measurement directions, leading to the corresponding collapse of the
wave function. But our initial question is not whether each individual
photon is a deterministic system, but whether we can define an isolated
\emph{enlarged} system, $E$, consisting of this measured particle,
the entire measurement device, and perhaps some environment interacting
with both of them, that behaves in a Laplacian deterministic way,
at least during a significant period of time. In the next Section,
we consider this question in detail.

\section{Assuming Laplacian determinism for the enlarged system}

Before entering into the main subject of the paper, let us make some
terminology precisions: Lately, some general consensus seems to have
been reached on the specific meaning of the two current terms \textquotedbl{}local
realism\textquotedbl{} and \textquotedbl{}determinism\textquotedbl{}
(see \cite{PhysRevA.87.052115}, for example): The first term ``demands
that the probability for obtaining outcomes $A$ and $B$ under settings
$a$ and $b$ can be written as a convex combination of products of
probabilities which depend only on the local setting and a shared
(hidden) variable''; the second term is a particular case of the
first one ``where the outcome probabilities are either $0$ or $1$''.
In the present paper, \textquotedbl{}determinism\textquotedbl{} will
mean what is meant by this term in Newtonian mechanics, and by well
known historical reasons we will call it here \textquotedbl{}Laplacian
determinism\textquotedbl{}. Obviously, its meaning does not coincide
with the previous definition of \textquotedbl{}determinism\textquotedbl{}.
Then, hereafter, in order to avoid any confusion between both concepts,
we call the last one \textquotedbl{}determinism as such\textquotedbl{}.
Let us be more specific:

Consider the time evolution of an isolated physical system, either
classical or quantum. Does it always exist, as a matter of principle,
a trajectory which, from some initial conditions, gives the values
of the different quantities of the system during some finite duration
of time, as it is, for example, the case in Newtonian mechanics, modulo
some regularity conditions? This constitutes the concept of Laplacian
determinism in nature. As we have already commented, this concept
is sensibly different from the concepts of local realism, and determinism
as such. More precisely, local realism and determinism as such, postulate
the existence of some hidden variables values behind the outcome of
a performed measurement, \emph{without requiring that these hidden
values remain the same after performing the measurement}. On the contrary,
Laplacian determinism (hereafter LD) for the isolated physical system
postulates the existence of the same \emph{standing} hidden variable
values (the unknown initial conditions) behind \emph{all} the successive
self-responses of the system along a certain finite time. In the present
paper, this isolated system will be the above \emph{enlarged system}
$E$

As a matter of fact, LD can be present under some circumstances, but
the problem we want to address is whether LD is \emph{always} present
\emph{regardless of our capacity of prediction in practice}. Therefore,
to discard LD it suffices to find at least one example where it is
contradicted by experimental data. This is what we are going to discuss
along the present paper. Our claim is that this kind of determinism
fails because it enters in contradiction with QM and, maybe, with
experiments (modulo some loopholes, also present in the well known
case where QM enters in contradiction with local realism).

QM is described in standard textbooks as an extremely successful non-deterministic
theory, since the result of measurements can only be predicted in
an statistical way. This is so even if the equation of motion (the
Schrödinger equation, for non relativistic QM) provides the future
state of the system when the initial conditions are specified. In
this sense, the time evolution of QM might be considered as deterministic.
It is the measurement process that introduces the non deterministic
nature of QM, since the result of a measurement can not, in general,
be predicted with certainty from the knowledge of the corresponding
quantum state. Moreover, this state is modified by the measurement,
leading to the well known collapse of the wave function. In other
words, the measurement process is invasive (in the specific sense
described in the Introduction) with respect to the system to be measured.

Thus, even if we admit that the quantum time evolution can be considered
as deterministic, the action of the measurement device on the system
under consideration appears as an extra ingredient that modifies the
conditions of that system. Once we accept this, the question arises:
Is it possible to postulate a Laplacian deterministic evolution that
describes, not only the quantum system (represented by $Q$) under
study, but also the measurement apparatus and, if necessary, the surrounding
environment? We have referred to this system as the \textit{enlarged
system} $E$. Our Laplacian deterministic hypothesis will be formulated
for $E$, and not for the quantum system, $Q$, to be measured. This
has to be clearly stated in order to avoid further misunderstandings:
The state of $Q$ may be (and in general will be) altered upon interaction
with the measurement device in a way that the theory seems unable
to predict. However, system $E$ as a whole will follow a predetermined
trajectory that evolves from some initial conditions, according to
the Laplacian deterministic hypothesis we want to be tested by confrontation
with the experiment.

The definition of determinism we introduced above, Laplacian determinism,
differs from local realism or determinism as such, can be discarded
by proposed experiments, and is based on testing some Leggett-Garg
inequalities, which take the form (see next Section)

\begin{equation}
|P(a,b)-P(a,c)|\le1-P(b,c),\label{inequalityintro}
\end{equation}
where $P(a,b)$ is the expected value associated to consecutive measurements
along polarization directions $a$ and $b$ (similarly for the rest
of magnitudes on this equation), on the \textbf{same photon}.

As shown in the next Section, we can find a scenario in which LD for
the enlarged system, $E$, enters in contradiction with QM. Similarly
to what happens with Bell inequalities, where local realism is falsified,
modulo two well known loopholes (see next), LD enters in contradiction
with QM, modulo the same loopholes. As already mentioned, this enlarged
system includes, at least, the measurement device besides the measured
system $Q$.

Similar results to the ones reported above have been previously obtained
by De Zela \cite{PhysRevA.76.042119} by adding to the deterministic
postulate a \emph{non contextuality} condition (\emph{freedom of choice,}
in the parlance of some authors), which amounts to assuming that initial
conditions and measurement directions are uncorrelated. Nevertheless,
this postulate needs to be justified, specially when LD is assumed,
since then these directions depend on the initial conditions. In the
present paper, we propose a kind of experiment for which LD by itself
would not entail contextuality, modulo the mentioned loopholes.

At the end of the paper we sketch the main differences between our
paper and De Zela's one \cite{PhysRevA.76.042119}.

\section{Laplacian determinism and the violation of the Leggett-Garg inequalities}

We start with a source of individual free photons that will provide
a large ensemble of photons numbered by $n=1,2,3,\cdots N$, all of
them prepared on the same quantum state, and consider the following
ideal experiment: We first fix a set of three space directions given
by the unit 3-vectors $\vec{a}$, $\vec{b}$ and $\vec{c}$. Then,
on each photon we perform two successive polarization measurements
along two randomly selected directions out of this set: We prepare
one photon and make two consecutive measurements, then prepare the
next photon, reset time to zero and proceed in the same way, etc ...

These \textquotedbl{}free\textquotedbl{} photons could evolve interacting,
to some extent, with their environment in an uncontrollable way, and
will certainly interact with the measurement apparatus. The latter
includes the device that randomly selects the two measurement directions
out of the three initially fixed directions. Consider now the physical
system that includes the photons to be measured, the experimental
facility and the interacting environment. In accordance with the two
precedent sections, we refer to this system as the \textit{enlarged
system}, represented by $E$, and assume that such system can be considered
as an isolated system during the whole experiment.

We now define our notion of determinism following what we have already
established, that is, following what happens, for example, with Newtonian
determinism (where, aside from some ``pathological'' cases, the
initial position and velocity, i. e., the initial conditions, allow
us to know some piece of the trajectory of a particle). We will make
the corresponding deterministic hypothesis, that we have called Laplace
determinism (LD), \textbf{for the enlarged system} $E$. Then the
successive polarization measurement outcomes, $\pm\hbar$, are fixed
from the initial conditions on $E$. Notice that we do not impose
any restrictions on the assumed initial conditions: In particular,
they could be non local, i.e. they could range over non causally connected
space-time regions.
\begin{figure}
\includegraphics[width=10cm]{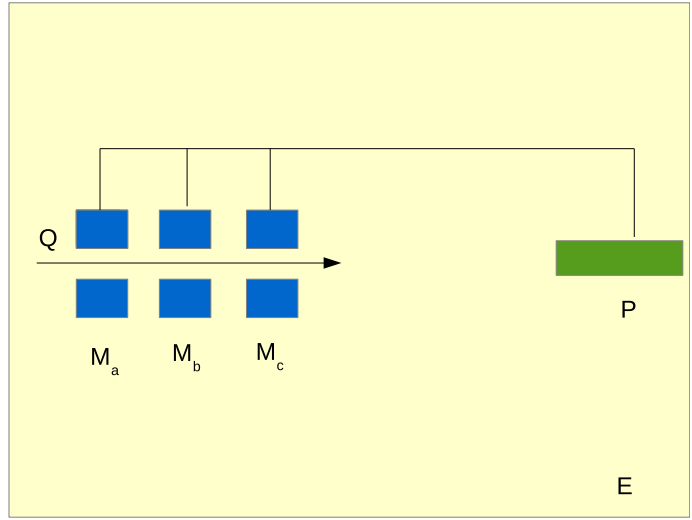}

\caption{A sketch of the measurement apparatus, as described in the text. $M_{a}$,
$M_{b}$ and $M_{c}$ perform polarization measurements on the system
$Q$ (an individual photon) according to the selection made by the
pseudorandom generator device $P$. Only two consecutive measurements
are performed. The experimental setup is contained on some enlarged
system $E$, which is assumed to be isolated.}

\end{figure}

Let us be more precise about our enlarged system and the measurement
process, which is sketched in Fig. 1. We denote by $\lambda$ the
initial conditions that we postulate to exist, complementary to the
prepared state of the quantum description, but leading to the same
statistical predictions than this quantum description. These initial
conditions will not belong exclusively to the photon, but to the whole
system $E$. Imagine that, each time we prepare a photon, system $E$
starts from different initial conditions, i. e., different $\lambda$
values. We will perform two consecutive polarization measurements
on each prepared photon. These two consecutive polarization measurements
will be performed at two randomly selected times, out of three fixed
values $t_{1},t_{2}$, and $t_{3}$, relative to the preparation time
(which is always reset to zero). To each selected time, $t_{1}$,
$t_{2}$ or $t_{3}$, we associate a constant measurement direction,
$\vec{a}$, $\vec{b}$ and $\vec{c}$, respectively. In other words,
we originally establish a given correspondence $\{t_{1}\rightarrow\vec{a},t_{2}\rightarrow\vec{b},t_{3}\rightarrow\vec{c}\}$
and keep it unchanged during the measurement process for the entire
set of photons. Thus, the randomness in the direction selection arises
only as a consequence of the randomness in the selected pair of times
out of the set $\{t_{1},t_{2},t_{3}\}$. The device (hereafter referred
to as $P$) that selects the pair of times, $(t_{1},t_{2})$, $(t_{1},t_{3})$
or $(t_{2},t_{3})$ (and so the choice of the pairs of measurement
directions) consists on a pseudo-random generator.

Then, let us denote by $S$ the values of the photon polarization
measurement outcomes, which are conveniently normalized to $\pm1$.
According to the LD postulate for our enlarged system, there exists
a function (unknown to us) which provides those outcomes for each
value of time $t_{i}$, starting from the initial conditions, i. e.,
the parameter values $\lambda$. Let us represent this function by
$S=S(\lambda,t_{i},\vec{x}(t_{i}))$, $i=1,2,3$ , with $\vec{x}(t_{i})\in\{\vec{a},\vec{b},\vec{c}\}$.
Notice that this notation for the function $S$ is actually redundant:
According to the above discussion, once we have fixed $\lambda$ and
$t_{i}$, the value of $S$ becomes determined, therefore we could
drop the argument $\vec{x}$ in $S$, although we will keep it for
convenience in the following discussion.

We now formally follow the original proof of usual Bell inequalities
\cite{Bell:1964kc} in order to arrive to the Leggett-Garg inequalities
(\ref{inequalityintro}) for our consecutive measurement outcomes.
Let us consider the following three expectation values:

\begin{equation}
P(a,b)=\int d\lambda\rho(\lambda)S(\lambda,t_{1},\vec{a})S(\lambda,t_{2},\vec{b}),\label{expectvaluab}
\end{equation}

\begin{equation}
P(a,c)=\int d\lambda\rho(\lambda)S(\lambda,t_{1},\vec{a})S(\lambda,t_{3},\vec{c}),\label{expectvaluac}
\end{equation}

\begin{equation}
P(b,c)=\int d\lambda\rho(\lambda)S(\lambda,t_{2},\vec{b})S(\lambda,t_{3},\vec{c}),\label{expectvalubc}
\end{equation}
where $\rho(\lambda)$ stands for the probability distribution of
the $\lambda$ values, which satisfies $\int d\lambda\rho(\lambda)=1$.
There are some subtleties related to the derivation of the above equations.
To begin with, the preparation of each photon, that is, the determination
of the corresponding initial conditions $\lambda$, might be correlated
with the chosen pair of directions corresponding to the two consecutive
measurements. In such a case, in (\ref{expectvaluab})-(\ref{expectvalubc}),
we should write $\rho_{ab}(\lambda)$, $\rho_{ac}(\lambda)$, or $\rho_{bc}(\lambda)$,
respectively instead of a common $\rho(\lambda)$, which will impede
us to complete the proof that follows below for the Leggett-Garg inequalities.
To avoid this, we assume that the times (and so the corresponding
measurement directions) selected by the device $P$ in Fig. 1, and
the events corresponding to the photon production, are space-like
separated (in a similar way to the one described in \cite{Scheidl16112010}).
Thus, we will assume that there is no such a correlation, that is,
we will assume freedom of choice.

However, there are still two loopholes, which are sometimes referred
to as \textquotedbl{}superdeterminism\textquotedbl{} (or in a equivalent
way \textquotedbl{}superrealism\textquotedbl{}) and \textquotedbl{}supercorrelation\textquotedbl{}
\cite{Scheidl16112010,raey}, also present in the proof of all kinds
of Bell inequalities. We will consider these loopholes in the next
Section, as well as a subtlety on the question of statistical \textquotedbl{}reproducibility\textquotedbl{}
which is specific to the present case, where we consider an enlarged
system $E$ which might have an enormous number of degrees of freedom.

Thus, we start from our expressions (\ref{expectvaluab})-(\ref{expectvalubc})
and go on with the derivation of our Leggett-Garg inequalities. We
take the difference

\begin{align}
P(a,b)-P(a,c) & =\int d\lambda\rho(\lambda)S(\lambda,t_{1},\vec{a})\label{difference}\\
\times & [S(\lambda,t_{2},\vec{b})-S(\lambda,t_{3},\vec{c})].
\end{align}

Henceforth, the proof of the inequalities goes formally along the
same lines as the proof of the original Bell inequalities in Bell's
seminal paper \cite{Bell:1964kc}. First, since $S^{2}(\lambda,t_{2},\vec{b})=1$,
the above difference can be written as

\begin{eqnarray}
P(a,b)-P(a,c) & = & \int d\lambda\,\rho(\lambda)S(\lambda,t_{1},\vec{a})S(\lambda,t_{2},\vec{b})\nonumber \\
 &  & \,\,\,[1-S(\lambda,t_{2},\vec{b})S(\lambda,t_{3},\vec{c})].\label{difference2}
\end{eqnarray}

Then, taking absolute values, we are led to

\begin{equation}
|P(a,b)-P(a,c)|\le\int d\lambda\rho(\lambda)[1-S(\lambda,t_{2},\vec{b})S(\lambda,t_{3},\vec{c})],\label{inequality}
\end{equation}
that is, to the well known Leggett-Garg inequality

\begin{equation}
|P(a,b)-P(a,c)|\le1-P(b,c),\label{inequality2}
\end{equation}
referring to two consecutive measurements on the \textbf{same photon}.

In QM, leaving aside the experimental difficulties to perform the
kind of experiment we are considering, the three mean values in (\ref{inequality2})
can be theoretically calculated as the corresponding expected values.
These values become $P(a,b)=\cos\,2\theta_{ab}$, where $\theta_{ab}$
is the angle between $\vec{a}$ and $\vec{b}$, irrespective of the
photon state prior to the first measurement and similarly for $P(b,c)$
and $P(a,c)$. Thus, inequality (\ref{inequality2}) becomes 
\begin{equation}
|\cos\,2\theta_{ab}-\cos\,2\theta_{ac}|+\cos\,2\theta_{bc}\le1,\label{inequality3}
\end{equation}
which is violated for example for $\theta_{ab}=\theta_{bc}=\pi/6$
and $\theta_{ac}=\pi/3$, in which case the left hand side of inequality
(\ref{inequality3}) reaches the value $3/2$.

Thus, for the \emph{enlarged system} consisting on our photons and
the measurement apparatus, including the device selecting the time
pairs and the corresponding measurement directions, plus the affecting
environment if any, the assumed LD enters in contradiction with quantum
mechanics. In other words, the essence of the present paper is the
following: could we consider \textbf{any system} $E$ larger than
the particle under study, so as to encompass the measurement device,
perhaps the laboratory and even beyond, in order to attain an isolated
enlarged system whose evolution fulfills the Laplacian deterministic
assumption? Our claim is that, if the appropriate measurements were
performed and inequality (\ref{inequality2}) was found to be violated,
as we expect from QM, then (up to two loopholes), no matter how large
$E$ was assumed to be, the answer to this question would be negative.

At this point, it is interesting to compare our result with a similar
one stated in \cite{0034-4885-71-2-022001}: There, under the following
three postulates, \emph{macroscopic realism per se}, \emph{noninvasive
measurability} and \emph{induction}, Leggett derives some Clauser-Horn-Shimony-Holt
(CHSH) inequalities \cite{PhysRevLett.23.880,Bellproced}, relating
the successive outcomes of a system with random dichotomic responses
against four types of measurements. Our LD postulate entails macroscopic
realism \emph{per se} and the induction postulate. Thus, a difference
between Leggett-Garg's approach and ours is that we replace the non-invasive
measurability for the measured system $Q$ by a Laplacian deterministic
postulate for the isolated enlarged system, which in our opinion is
full of epistemological meaning. Since this determinism is assumed
for the whole system $E$, this implies that all quantities in this
system, including the measuring device, are defined at all times once
some starting initial conditions $\lambda$ belonging to $E$ are
specified (even if unknown). Furthermore, since the measured system,
$Q$, is a part of $E$ and $E$ is Laplacian deterministic, so is
the $Q$ system, although its corresponding initial conditions, $\lambda$,
do not belong exclusively to $Q$, but to $E$. This reasoning \emph{formally}
converts $Q$ into a non-invasive measured system, at the cost of
allowing for initial conditions not entirely belonging to $Q$.

\section{Two loopholes and a subtle question}

Because of the assumed space-like separation of the two events, particle
preparation and choice of the measurement directions, we have discarded
any kind of correlation between the initial conditions $\lambda$,
on the one hand, and the selected pair of measurement directions $(\vec{a},\vec{b})$,
$(\vec{a},\vec{c})$, or $(\vec{b},\vec{c})$, on the other.

As pointed out above, there is, however, a first loophole against
this assumed no correlation, the so called \textquotedbl{}superdeterminism\textquotedbl{}
(also called \textquotedbl{}superrealism\textquotedbl{}) (see, for
example, \cite{Scheidl16112010,raey}): Let us consider the past light
cone associated to the particle preparation that fixes the corresponding
initial conditions $\lambda$, and also the two past light cones associated
to selecting the first and second directions, say $\vec{a}$ and $\vec{b}$.
The $\lambda$ past cone and the $\vec{a}$ cone will share a common
region, and similarly for the $\lambda$ and $\vec{b}$ cones. Then,
these common regions could be the source of some correlation between
$\lambda$ and $(\vec{a},\vec{b})$.

The second loophole or \textquotedbl{}supercorrelation\textquotedbl{}
\cite{raey} goes this way: without resorting to any shared region
among the backward cones of the \textquotedbl{}superdeterminism\textquotedbl{},
some correlations between initial conditions and measurements settings
could be originally arranged between space-like separated places on
the basis of what Bell \cite{refId0} called a possible conspiratory
arrangement in Nature.

As we have already commented in the precedent Section, both loopholes
are also present in the proofs of the different versions of Bell's
inequalities. In \cite{Colbeck2012,Koh2012}, the question of these
two loopholes has been addressed in the perspective of a procedure
to reduce a given, small enough, degree of correlation to a virtually
absence of correlation, assuming the non-signaling postulate, a procedure
that we cannot apply here since this postulate is not satisfied in
the case of our measured system $Q$, i.e., in the case of the considered
photons.

To end the present Section we refer to a subtle difficulty: perhaps
the different initial conditions $\lambda$ never repeat themselves
completely. This should not be so strange in the present case of an
enlarged macroscopic system whose number of degrees of freedom might
be huge. Because of this hypothetical non repetition, we could have
three hypothetical subsets, let us designate them, in an evident notation,
$\{\lambda_{ab}\}$, $\{\lambda_{ac}\}$ and $\{\lambda_{bc}\}$,
that might never achieve a standing equality $\{\lambda_{ab}\}=\{\lambda_{ac}\}=\{\lambda_{bc}\}\equiv\{\lambda\}$.
The lack of this equality would impede the proof of the Leggett-Garg
inequalities that we have displayed in the previous Section. Nevertheless,
this hypothetical inequality is not the problem, since what is physically
relevant is not the, for example, $\lambda_{ab}$ values by themselves,
but the corresponding function values $S(\lambda_{ab},t_{1})$, entering
in the expected value $P(a,b)$, in (\ref{inequality2}). But in the
present case with photons, we can relay on the statistical ''reproducibility''
of the measurement outcomes, meaning by this that, in accordance to
QM, those expectation values become stabilized for a sufficiently
large number, $N$, of measurement pairs. This implies that, beyond
$N$, we still could formally write $P(a,b)$ in (\ref{expectvaluab})
using the same finite subset, $\{\lambda_{ab}\}$, that we have used
just for $N$. Thus, in practice, because this ``reproducibility''
property, (which, in the present case, because of the quantum predictions,
can be taken as an experimental fact) we can use a finite number of
$\lambda_{ab}$ values, and the same for the $\lambda_{ac}$ and the
$\lambda_{bc}$, to avoid our initial difficulty.

\section*{Conclusions}

Let us summarize the hypothesis we introduced in order to derive our
Legget-Garg inequalities (\ref{inequality2}), based on LD postulate
for the enlarged system, as compared to other hypothesis currently
made in similar areas: 
\begin{itemize}
\item We assume \textit{Laplace determinism} defined as follows: One can
define an enlarged isolated system $E$ that includes the particles
to be measured, the experimental setup and the surrounding environment,
such that all variables in that system evolve predictively from some
unknown initial conditions $\lambda$ belonging entirely to this isolated
system. This amounts to consider the measured system as another Laplacian
deterministic system, one whose corresponding initial conditions do
not belong entirely to it, but to the isolated enlarged system. LD
implies induction: Initial conditions are not affected by measurements
performed later in time. 
\item We assume that there is no correlation between the initial conditions
$\lambda$, on the one hand, and measurements directions, on the other
hand, by arranging the experimental setting such that particle preparation
and the corresponding two measurement directions choice are space-like
separated events. 
\item We need to use statistical \textit{reproducibility,} which in the
case of photons is predicted by QM, and thus could be considered as
an experimental fact. 
\end{itemize}
Leaving the particular case of photons, and going to the general case
of a measured system, macroscopic or not, whose measured outcomes
are dichotomic (say, for instance, $\pm1$) and apparently random,
it is straightforward to see that we still could derive some Leggett-Garg
inequalities by postulating LD for the corresponding enlarged system,
provided that we experimentally guarantee a space-like separation
between system preparation and selection setting, and provided that
the measurement outcomes obey statistical \textit{reproducibility}.

Obviously, in the general case, the two loopholes, ``superdeterminism''
and ``supercorrelation'' remain open, although one could help to
partially close them by resorting to the procedure of free randomness
amplification explained in \cite{Colbeck2012}, provided that the
non-signaling postulate be actually satisfied in practice.

In all, modulo these two loopholes, the meaning of the attained violation
of Leggett-Garg inequalities (\ref{inequality2}) is the impossibility
of finding any enlarged system, $E$, whatever large it be, that allow
us to describe the photon polarization measurement outcomes as belonging
to this system $E$ obeying LD. In other words, this means the violation
of LD in Nature.

Then, before finishing, let us sketch which are the main differences
between our paper and the preceding one from De Zela \cite{PhysRevA.76.042119}:

First: We have made his non contextuality assumption (freedom of choice
assumption) more consistent with the central postulate of LD by virtually
adopting an experimental setting where preparation of the system and
setting choice are space-like separated events.

Second: In a case like the present one, and also like the one by De
Zela, where the considered enlarged system has a huge number of degrees
of freedom, we have shown how unavoidable is to rely on the statistical
\textquotedbl{}reproducibility\textquotedbl{} in order to prove Leggett-Garg
inequalities from LD.

Third: We have clearly identified the two loopholes left, \textquotedbl{}superdeterminism\textquotedbl{}
and \textquotedbl{}supercorrelation\textquotedbl{}, and in the general
case, that is, apart from the particular case considered in the present
paper, we point out the possibility of approaching their closeness
by using a protocol to reduce the scope of these two loopholes \cite{Colbeck2012},
provided the non-signaling postulate be actually satisfied.

Finally: contrarily to De Zela, we have not needed to treat separately
the two cases of non-invasive or invasive measurability since our
central postulate, LD, for the enlarged system, $E$, entails by itself
LD (and so \emph{formal} non-invasiveness) for the measured system
$Q$, although the $Q$ initial conditions do not exclusively belong
to $Q$ itself.
\begin{acknowledgments}
This work has been supported by the Spanish Ministerio de Educación
e Innovación, MICIN-FEDER project No. FIS2012-33582, and by Projects
FPA2011-23897 and ``Generalitat Valenciana'' grant PROMETEO/2009/128.
Useful discussions with J. Kofler and E. Roldán are gratefully recognized.
\bibliographystyle{plain}
\bibliography{Bell}

\begin{thebibliography}{13}
\expandafter\ifx\csname natexlab\endcsname\relax\def\natexlab#1{#1}\fi
\expandafter\ifx\csname bibnamefont\endcsname\relax
  \def\bibnamefont#1{#1}\fi
\expandafter\ifx\csname bibfnamefont\endcsname\relax
  \def\bibfnamefont#1{#1}\fi
\expandafter\ifx\csname citenamefont\endcsname\relax
  \def\citenamefont#1{#1}\fi
\expandafter\ifx\csname url\endcsname\relax
  \def\url#1{\texttt{#1}}\fi
\expandafter\ifx\csname urlprefix\endcsname\relax\def\urlprefix{URL }\fi
\providecommand{\bibinfo}[2]{#2}
\providecommand{\eprint}[2][]{\url{#2}}

\bibitem[{\citenamefont{Leggett}(2002)}]{0953-8984-14-15-201}
\bibinfo{author}{\bibfnamefont{A.~J.} \bibnamefont{Leggett}},
  \bibinfo{journal}{Journal of Physics: Condensed Matter}
  \textbf{\bibinfo{volume}{14}}, \bibinfo{pages}{R415} (\bibinfo{year}{2002}),
  \urlprefix\url{http://stacks.iop.org/0953-8984/14/i=15/a=201}.

\bibitem[{\citenamefont{Leggett and Garg}(1985)}]{PhysRevLett.54.857}
\bibinfo{author}{\bibfnamefont{A.~J.} \bibnamefont{Leggett}} \bibnamefont{and}
  \bibinfo{author}{\bibfnamefont{A.}~\bibnamefont{Garg}},
  \bibinfo{journal}{Phys. Rev. Lett.} \textbf{\bibinfo{volume}{54}},
  \bibinfo{pages}{857} (\bibinfo{year}{1985}).

\bibitem[{\citenamefont{Kofler and Brukner}(2013)}]{PhysRevA.87.052115}
\bibinfo{author}{\bibfnamefont{J.}~\bibnamefont{Kofler}} \bibnamefont{and}
  \bibinfo{author}{\bibfnamefont{i.~c.~v.} \bibnamefont{Brukner}},
  \bibinfo{journal}{Phys. Rev. A} \textbf{\bibinfo{volume}{87}},
  \bibinfo{pages}{052115} (\bibinfo{year}{2013}),
  \urlprefix\url{http://link.aps.org/doi/10.1103/PhysRevA.87.052115}.

\bibitem[{\citenamefont{De~Zela}(2007)}]{PhysRevA.76.042119}
\bibinfo{author}{\bibfnamefont{F.}~\bibnamefont{De~Zela}},
  \bibinfo{journal}{Phys. Rev. A} \textbf{\bibinfo{volume}{76}},
  \bibinfo{pages}{042119} (\bibinfo{year}{2007}).

\bibitem[{\citenamefont{Bell}(1964)}]{Bell:1964kc}
\bibinfo{author}{\bibfnamefont{J.~S.} \bibnamefont{Bell}},
  \bibinfo{journal}{Physics} \textbf{\bibinfo{volume}{1}}, \bibinfo{pages}{195}
  (\bibinfo{year}{1964}).

\bibitem[{\citenamefont{Scheidl et~al.}(2010)\citenamefont{Scheidl, Ursin,
  Kofler, Ramelow, Ma, Herbst, Ratschbacher, Fedrizzi, Langford, Jennewein
  et~al.}}]{Scheidl16112010}
\bibinfo{author}{\bibfnamefont{T.}~\bibnamefont{Scheidl}},
  \bibinfo{author}{\bibfnamefont{R.}~\bibnamefont{Ursin}},
  \bibinfo{author}{\bibfnamefont{J.}~\bibnamefont{Kofler}},
  \bibinfo{author}{\bibfnamefont{S.}~\bibnamefont{Ramelow}},
  \bibinfo{author}{\bibfnamefont{X.-S.} \bibnamefont{Ma}},
  \bibinfo{author}{\bibfnamefont{T.}~\bibnamefont{Herbst}},
  \bibinfo{author}{\bibfnamefont{L.}~\bibnamefont{Ratschbacher}},
  \bibinfo{author}{\bibfnamefont{A.}~\bibnamefont{Fedrizzi}},
  \bibinfo{author}{\bibfnamefont{N.~K.} \bibnamefont{Langford}},
  \bibinfo{author}{\bibfnamefont{T.}~\bibnamefont{Jennewein}},
  \bibnamefont{et~al.}, \bibinfo{journal}{Proceedings of the National Academy
  of Sciences} \textbf{\bibinfo{volume}{107}}, \bibinfo{pages}{19708}
  (\bibinfo{year}{2010}),
  \eprint{http://www.pnas.org/content/107/46/19708.full.pdf+html},
  \urlprefix\url{http://www.pnas.org/content/107/46/19708.abstract}.

\bibitem[{\citenamefont{Vervoort}(2013)}]{raey}
\bibinfo{author}{\bibfnamefont{L.}~\bibnamefont{Vervoort}},
  \bibinfo{journal}{Foundations of Physics} \textbf{\bibinfo{volume}{43}},
  \bibinfo{pages}{769} (\bibinfo{year}{2013}), ISSN \bibinfo{issn}{0015-9018},
  \urlprefix\url{http://dx.doi.org/10.1007/s10701-013-9715-7}.

\bibitem[{\citenamefont{Leggett}(2008)}]{0034-4885-71-2-022001}
\bibinfo{author}{\bibfnamefont{A.~J.} \bibnamefont{Leggett}},
  \bibinfo{journal}{Reports on Progress in Physics}
  \textbf{\bibinfo{volume}{71}}, \bibinfo{pages}{022001}
  (\bibinfo{year}{2008}).

\bibitem[{\citenamefont{Clauser et~al.}(1969)\citenamefont{Clauser, Horne,
  Shimony, and Holt}}]{PhysRevLett.23.880}
\bibinfo{author}{\bibfnamefont{J.~F.} \bibnamefont{Clauser}},
  \bibinfo{author}{\bibfnamefont{M.~A.} \bibnamefont{Horne}},
  \bibinfo{author}{\bibfnamefont{A.}~\bibnamefont{Shimony}}, \bibnamefont{and}
  \bibinfo{author}{\bibfnamefont{R.~A.} \bibnamefont{Holt}},
  \bibinfo{journal}{Phys. Rev. Lett.} \textbf{\bibinfo{volume}{23}},
  \bibinfo{pages}{880} (\bibinfo{year}{1969}).

\bibitem[{\citenamefont{Bell}(1971)}]{Bellproced}
\bibinfo{author}{\bibfnamefont{J.~S.} \bibnamefont{Bell}}, in
  \emph{\bibinfo{booktitle}{Proceedings of the international School of Physics
  "Enrico Fermi", course IL}} (\bibinfo{publisher}{Academic},
  \bibinfo{year}{1971}), p. \bibinfo{pages}{171}.

\bibitem[{\citenamefont{{Bell, J. S.}}(1981)}]{refId0}
\bibinfo{author}{\bibnamefont{{Bell, J. S.}}}, \bibinfo{journal}{J. Phys.
  Colloques} \textbf{\bibinfo{volume}{42}}, \bibinfo{pages}{C2}
  (\bibinfo{year}{1981}).

\bibitem[{\citenamefont{Colbeck and Renner}(2012)}]{Colbeck2012}
\bibinfo{author}{\bibfnamefont{R.}~\bibnamefont{Colbeck}} \bibnamefont{and}
  \bibinfo{author}{\bibfnamefont{R.}~\bibnamefont{Renner}},
  \bibinfo{journal}{Nat Phys} \textbf{\bibinfo{volume}{8}},
  \bibinfo{pages}{450} (\bibinfo{year}{2012}), ISSN \bibinfo{issn}{1745-2473},
  \urlprefix\url{http://dx.doi.org/10.1038/nphys2300}.

\bibitem[{\citenamefont{Koh et~al.}(2012)\citenamefont{Koh, Hall, Setiawan,
  Pope, Marletto, Kay, Scarani, and Ekert}}]{Koh2012}
\bibinfo{author}{\bibfnamefont{D.~E.} \bibnamefont{Koh}},
  \bibinfo{author}{\bibfnamefont{M.~J.~W.} \bibnamefont{Hall}},
  \bibinfo{author}{\bibfnamefont{M.~J.~W.} \bibnamefont{Setiawan}},
  \bibinfo{author}{\bibfnamefont{J.~E.} \bibnamefont{Pope}},
  \bibinfo{author}{\bibfnamefont{C.}~\bibnamefont{Marletto}},
  \bibinfo{author}{\bibfnamefont{A.}~\bibnamefont{Kay}},
  \bibinfo{author}{\bibfnamefont{V.}~\bibnamefont{Scarani}}, \bibnamefont{and}
  \bibinfo{author}{\bibfnamefont{A.}~\bibnamefont{Ekert}},
  \bibinfo{journal}{Phys. Rev. Lett.} \textbf{\bibinfo{volume}{109,}},
  \bibinfo{pages}{160404} (\bibinfo{year}{2012}), \eprint{1202.3571}.

\end{thebibliography}
 \end{acknowledgments}

\end{document}